\documentclass[12pt]{article}
\usepackage{amsmath}
\usepackage{amssymb}
\renewcommand{\title}[1]{%
    \bigskip%
    \begin{center}%
    \Large\bf #1%
    \end{center}%
    \vskip .2in}

\renewcommand{\author}[1]{
   {\begin{center}
    #1
    \end{center}}}
\newcommand{\address}[1]{\vspace{-1.7em}\vspace{0pt}
    {\begin{center}
    \it #1
    \end{center}}}

\begin{document}
\title{\bf{ A new scalar electrodynamics for fracton gauge theory }}

\author
{  
Sk. Moinuddin   $\,^{\rm a, b}$, 
Pradip Mukherjee  $\,^{\rm a,c}$}

\address{$^{\rm a}$Department of Physics, Barasat Government College,Barasat, India}
\address{$^{\rm b}$\tt
dantary95@gmail.com  }
\address{$^{\rm c}$\tt mukhpradip@gmail.com}

\abstract{A new representation  of the scalar electrodynamics is discovered which gives a more redundant description of electromagnetic theory  and
 suitable to construct an appropriate matter action which contains  two global symmetries . The symmetries of the  model when localized by gauge principle , provides a  co-variant  fracton gauge theory. }

\section{Introduction}

 ``When I first heard about fractons, I said there is no way this could be true, because it completely defies my prejudice of how systems behave,” - such was the reaction of Nathan Seiberg, one of the most important physicists of our time, about fractons \cite{ns}.  No doubt   he did not stop at this, but made useful contributions  to the development of fracton phase research \cite{sei}. Similarly, another promising physicist and a leader of the research on the fracton gauge theory ,  Michael Pretko, in a recent review  \cite{pret1}, observed -- ``It remains an open problem whether this theory can be rewritten in a manifestly rotationally invariant way"  .   These reactions  are not surprising as fractons are the most exotic among  the quasi particles — particle-like entities that emerge out of complicated interactions between many strongly interaccting electrons  inside a material \cite{ash}. But fractons are bizarre even compared to other exotic quasi-particles, because they are totally immobile or able to move only in a limited way.

   The fracton phase of matter is thus, one of the marvels of strongly correlated electronic motion. Discovered only in the first decade of this century, it has already made its place in the challenge-list of theoretical physics. Many important research has been conducted on both gapped and gappless fracton phases of matter \cite{pret1}\cite{nand}. It can be said with confidence that the theory of fractons has reached a mature stage. Though an isolated fracton has not been observed experimentally till date , it is believed from the behavior of the glassy materials that the imprint of fracton has been registered as far back as 2005  \cite{nand} \cite{cham}. But there are many gaps still remaining in the understanding  the physics of the fracton phase of matter. A prominent example is the   connection between the  gapped and the gapless phases. The    belief  of many theoreticians notwithstanding,    the theory connecting  the gapless phase to the gapped phase of fracton \cite{nand} is still in its infancy . So we will be confined to the gappless phase where the powerful  apparatus of field theory is applicable \cite{pret}.
   
   
   We see that the  fractons  are distinguished by their
 very low (often vanishing) mobility \cite { cham}  \cite{haa}  in contrast with the other (quasi) particles. Due to its immobility fractons are very slow to respond to an applied electric field . But composites of fractons  can move freely. On the theoretical side this immobility of the fracton in isolated state and mobility in presence of other fractons   is really a challenging phenomenon.   Note that  the immobility of fractons can be explained by  some special symmetries . The theorists introduced a  phenomenology based on a  tensor gauge theory \cite {pret1}. The tensor gauge theory is assumed to have dipole moment conservation,  in addition to the well known electric charge conservation. The conservation of dipole moment was introduced phenomenologically . As we have mentioned earlier ,
 the breakthrough \cite {pret} points to  another symmetry. The novelty of Pretko's theory consists in the correct understanding that one requires a gauge theory to generate a conservation law. The  algorithm of \cite {pret} involves a multiplet of  real scalar fields  $ (\phi_1,\hspace{.1cm} \phi_2, ... ) $ which form a representation of the gauge symmetry of the theory. A combination of the components were found  in such a way that it carries a  symmetry   $ \phi \to e^{i\alpha(x)}\phi  $ .  Thus a Lagrangian was constructed with the combinations as the constituents in a manner so that  space dependant  rotation symmetry is  ensured.Expanding in an a perturbation series, upto a 
  certain order of the space depended  rotation symmetry , the desired theory with two conserved charges was obtained. Clearly the algorithm is based on an improvisation rather than following from some basic premises. . The phenomenology of fracton matter was thus explained. Additionally, many interesting interactions of fractons were predicted by the theory in as diverse fields , from    elasticity to gravity \cite {pret1}.
  
   The theory mooted in \cite{pret} was thus accepted as the basis of fracton gauge theory and is being followed till date.

   However , in the grandeur of success some disturbing points were overlooked.
   
   \begin{enumerate}

\item The interaction between the fractons is electromagnetic in nature. This is an Abelian Lie group with a single gauge . A Hamiltonian analysis of the theory will generate only one primary first class constraints. The gauge generator is a one parameter operator \cite{blago} . How come two independent gauge symmetries may be generated here?

\item   The construction of \cite {pret} does not give any idea how the modified matter Lagrangian
originates. 

\item Perhaps the biggest shortcoming of the theory of \cite{pret} is the violation of the boost symmetry. This would make the gauge  symmetry meaningless. The gauge invariance valid in one inertial frame may be broken  in another reference frame \cite{bd}.

\end{enumerate}

 The gauge theory which we are  going to report here is a lineage of the existing fracton gauge theory. However,  the points of departures which are required is suggested by the above. Our theory is therefore constructed right from the beginning  in a covariant way. Also we chose the fracton matter in a way where the objections are  nonexistent. How this construction is realized will be  unfolded below.

 \hskip.5cm  The focus of our paper is to  construct an appropriate covariant charged matter field action which has two different global symmetries.  It would be convenient if we start with a list of the new ideas that have  been invoked here.  .
 
 \begin{enumerate}
 
 \item
 
  An U(1) gauge theory with usual Lagrangian structure , it is impossible to have more than one  independent gauge generator  \cite{hp}. Therefore we must consider more general Lagrangian theory that is the Lagrangian must be of higher order \cite{gt}. In the higher order theories it has been proved that the number of independent conservation laws may be greater than the number of independent gauge parameters \cite{pmb}\cite{pb}.

\hskip.5cm  {\bf{This is the first new idea which we invoke in our paper.}}
 
 \item
 
  The higher order Lagrangian theories are being studied for a long time \cite{ostro}, but the main problem is the appearances of instability (ostrogradsky  instability ) \cite{ostro}. So one may ask how such theories may be useful in the context of real physical calculation. {\bf{ Now comes the second new idea in our work . }} This may seem astounding that this new idea is borrowed from the theory of modified gravity in the context of cosmology. To bridge the violation of general relativity (GR) at very large distances and simultaneously holding of GR in the solar system distances Dvali-Gabadadze- Porrati introduce a model (DGP model)\cite{dva}. We are not concern about cosmology here , but just note in a certain limit of DGP model  the concept of a new neutral scalar matter theory was originated \cite{nic}. We can treat it as a new scalar matter , stripping it of all its gravitational origin. This scalar matter has a higher order Lagrangian with the equations of motion having no higher order term (such term cancel out as we will see below). This scalar is known as the  Galileons . The point is that the action is invariant under linear shift 
  
   \begin {equation}
  \pi \to \pi + (a  +b_{\mu} x^\mu) \label{0}  
    \end{equation} 
    
The symmetry  (\ref{0}) is a global symmetry since the transformation parameters are constants. But the elements of the symmetry group are dependent on position. This point should be borne in mind for subsequent application.

\hskip .5cm {\bf{ The second new idea which we invoke here is the Galileons are the basic units of constructing the charged fracton matter.}}

\end{enumerate}

\hskip .5cm How these ideas can be implemented in the context of the fracton gauge theory and how it complements the existing set of the arts to produce a covariant gauge theory ,  will be the content of our report.

 \hskip .5cm  After the introductory section we report the construction of the scalar theory which has both phase rotation symmetry and a linear complex shift symmetry \footnote{ Note that the realization of the complex shift symmetry is a noval idea in the literature.}.  The action was established  comprehensively from the various symmetry requirements almost uniquely. Thus we can provide  not only a `manifestly rotational  symmetry ' but  a co-variant one also. This is a testimony that  the theory presented  by us   reveals a deep connection with the physics of fracton  matter . It can be also appreciated that our action can easily be generalized for any type of interaction readily .  The reason is that our theory is  established  on  two cardinal concepts of theoretical physics, namely, covariance and gauge symmetry. This is remarkable because  a careful survey of the literature reveals that not many papers can be found  where a derivation from the first  principles is available. 
 
  \hskip.5cm In the next section a comprehensive analysis of the different symmetries have been discussed. The conserved charges are computed using Noether's theorem . The connection between the two symmetries came out automatically and the corresponding conserved charges are readily identified as the electric charge and dipole moment. Clearly the existing phenomenology of the fracton matter  followed spontaneously.

 \hskip.5cm From  section IV ,     the applications of the gauge principle to the phase rotation symmetry, complex shift symmetry and space time symmetries are computed explicitly . In this process we have resolved  various  puzzling aspects. The `gauging the symmetry approach' will be our only tool. We start with enunciating the precise algorithm \cite{Uti} in a short review  by applying it to the complex Klein Gordon (K-G) field. The rest of the section will be devoted to the localization of our model   i.e. the un-gauged fracton model. As a result of localization  in addition to the well known  Maxwell's theory with vector gauge field ,  we get another theory of  a tensor gauge field $B_{\mu\nu}$. Look how methodically the tensor gauge field appears in our analysis as a consequence of the symmetries. The properties of this tensor gauge field has so far been obtained from the fracton phenomenology. In the next section  we give our analysis of the behavior of  fracton in external electromagnetic field and show that the so called generalised Maxwell theory is another way of representing  the electromagnetic field.
 These results follow smoothly from our algorithm. Once again we see the efficacy of it .

\hskip.5cm In section  VI, we discuss  the  localization of the space-time symmetry (Poincare symmetry). A surprising fallout is the emergence of another second rank tensor gauge field. Note carefully that in the existing  theory the same tensor gauge field is assumed in both electromagnetic and gravitational sectors. It is understandable that our theory correctly represent the two interaction with two different  tensor gauge theories . It will indeed be very difficult to explain how a field can have both gravitational and electromagnetic  interaction.   As an extra dividend, we derived   the self interaction of the fracton , a puzzling issue in the present state of art. Note that though we are working in the backdrop of  the classical field theory , the results obtained in this paper are far more general, because our work relies on symmetries like Lie groups of symmetries and Poincare symmetries.

\section{A new scalar electrodynamics}

This section is one of the most important parts of our work because it introduces the construction of a new field theoretic representation of the $U(1)$ group which would go henceforth by the name`` new scalar electrodynamics "
. To understand  different intricacies of the model we  divided our discussion in a number of subsections. In the first subsection would illustrate the application of the gauge  principle to a  well known field theory.

\subsection{Application of the gauge principle illustrated}
As declared above we start with the  concept of gauge principle and how to use it by reviewing the deduction of the scalar electromagnetic theory i.e. to derive the well known   model  of
the usual scalar electrodynamics, which is given by 
\begin{eqnarray}
 {\mathcal{L}_{kg}} = \partial_\mu\phi  \partial^\mu\phi^{*}-J^{\mu}A_{\mu} -\frac{1}{4} F^{\mu\nu}F_{\mu\nu}\label{99.1}
\end{eqnarray}

 where $ F_{\mu\nu}=( \partial_{\mu}A_{\nu}-\partial_{\nu}A_{\mu} ) $ , $ A_{\mu}$ is the 4 potential and $ J^{\mu}$ is the conserved current;
 $ J^\mu = -i(\phi\partial^\mu\phi^* - \phi^*\partial_\mu \phi )$ .  
  A beautiful illustration of the interplay of different symmetries including the gauge symmetry is revealed in our way towards establishing  the Lagrangian (\ref{99.1}). The free Lagrangian ($\partial_\mu\phi \partial^\mu\phi^{*} $) representing the matter field has a global  symmetry 
  \begin{eqnarray}
 \phi & \to & e^{i\Lambda}\phi \nonumber\\
 \phi^* & \to & e^{-i\Lambda}\phi*
 \label{gauge}
 \end{eqnarray}
 
 Note that here  $\partial
 _{\mu}\phi $ transform in the identical way as the field $\phi$. Technically this is called covariance . This covariance of the gradient however is lost if we localize the symmetry that is making $\Lambda$ as function of space and time . Now this lost symmetry can be regained by defining a covariant derivative ($D_{\mu} \phi = \partial_{\mu}\phi + iA_{\mu}\phi$) and replacing the gradient everywhere by the covariant derivative. Here $A_{\mu}$ is a one form which transform appropriately so that $D_{\mu} \phi$ is covariant. $A_{\mu}$ is the gauge field which is identified with the electromagnetic potential. So using the gauge principle we get the modified Lagrangian as $ D_\mu \phi D^{\mu}\phi^{*}$. Simplification of this  gives the Lagrangian (\ref{99.1}) without the last term (dynamical term) that is the form of $\phi$ interacting with $A_{\mu}$. This procedure is well known and term as gauge principle. To apply the gauge principle the following points should be noted 
 
 \begin{enumerate}
\item
The free theory Lagrangian must have a global symmetry which becomes the symmetry of the interacting field on localization.

\item
 Matter interacts with the electromagnetic field through the charge. Also there is one independent gauge field as the symmetry group is a one parameter Lie group. Remember that the number of gauge symmetry is equal to number of first class constraint in phase space \cite{dir} .

 \end{enumerate}

\subsection {A new field theoretic representation for U(1) symmetry  }

 Now we turn back to our problem of interest, In the introductory remarks it has been mentioned that one of the most bizarre aspects is the existence of two conservation laws . This means that there are two primary first class constraints in the theory. But we know that the fundamental interaction here is  Max-well's electrodynamics  which is  U(1) gauge theory that is  one parameter  abelian group . From the Hamiltonian point of view this implies that there should be not more that one primary first class constraint. This again states that the number of independent conservation laws cannot exceed one. This apparent paradox actually lies in the representation chosen for the group. In scalar electrodynamics the   usual action contains a Lagrangian which may at-most depend on $\partial_{\mu}\phi$ . Thus the usual Lagrangian  is of first order, so that equation of motion are maximum second order. Naturally , there is no scope of such matter theory to have more than one independent conservation law. This paradox disappears if we allow higher order theories to enter in our scenario . We know that the relation between the number of independent conservation laws may be greater than the number of primary first class constraints. So the new scalar electrodynamics must have a matter sector which is a theory containing higher order derivatives in th Lagrangean.But then  one may legitimately argue that the higher order theories are known to have in general  higher order equations of motion. These are plagued with
  essential instability (ostrogradsky instability) \cite{gt}, \cite{ostro}. {\bf{So it appears that we are in a fix}}. 

 \hskip .5cm  The way-out from this fix is obtained surprisingly from a distant
 quarter. In the research on modified gravity in a certain  limit of  Dvali-Gabadadze- Porrati (DGP model there emerged a scalar ( Galileon scalar) theory which has ordinary second order equations of motion in-spite of having higher order action. In their  seminal paper \cite{nic}  Rattazzi at-all demonstrated that a very special combination of  first and second derivatives of the  scalar field  may serve as the corresponding Galileon  Lagrangian. In (3+1) dimension there are five Galileon actions
  $L_1 $to
  $L_5 $. Of them 
  $L_1 = \pi$ is trivial and 
  $L_2 = \partial_\mu\pi \partial^\mu\pi$ is the usual Lagrangian . The first nontrivial Gallileon Lagrangian is       
\begin{eqnarray}
 L_{3}=\Box\pi\partial_{\mu}\pi\partial^{\mu}\pi \label{l3}
 \end{eqnarray}  
  
 $L_4 $ and  $L_5 $ are the other nontrivial Galileon term. Among them $L_4 $ is given by \footnote{This action is not identical to the usual expression in \cite{nic} , but one can easily check that they are equivalent.}
 
 \begin{eqnarray}
 L_{4}=\bigg(\partial_{\mu}\pi\partial^{\mu}\partial_{\nu}\pi\partial^{\nu}\partial_{\rho}\pi\partial^{\rho}\pi  - \Box\pi\partial_{\nu}\pi\partial^{\rho}\pi \partial^{\nu}\partial_{\rho}\pi \bigg)\label{011}
 \end{eqnarray} 
 
  All the nontrivial Galileon Lagrangian satisfy the following properties :
\begin{enumerate}
\item
Despite being higher derivative theories, they are free of Ostrogradsky type ghosts, since the equation of motion following from Galileons action are second order.

\item
The Galileon  field is a scalar under Poincare transformations and  has a unique symmetry under the shift,
 
     \begin{equation} 
\pi \to \pi + a +b_{\mu} x^\mu
\end{equation}

  where a and $b_{\mu}$  are  real numbers. A common mistake should be avoided here. The symmetry transformation  given above is a global symmetry, in spite of its $ x $ dependence. Note  also that we some times indicate the first term $a$ as a translation and second part $ b_{\mu} x^\mu$  as shift while some time we will refer the whole transformation as shift. It will be understood from the context, which meaning is implied.
 
\end{enumerate} 

 Let us note that though the concept of Galileonic theories are very important for our purpose, there is no question of following  it top to bottom .  We have to build an action in terms of a complex scalars. But they have a correspondence. A shift of the Galileon space corresponds to a complex shift in the group space of the complex scalars.  So armed with the novel scalar field we may pass to the model building.
 
 \subsection{Building the model}

 \hskip .5cm  We can define a complex scalar field using the doublet of Galileon scalar as a basic elements. 
\begin{eqnarray}
\phi &=&\pi_{1}+i\pi_{2} \nonumber\\
\phi^{*}&=&\pi_{1}-i\pi_{2} \label{98}
\end{eqnarray} 
 where $\pi_{1}$ and $\pi_{2}$  are Galileon scalars . $\phi$ and $\phi^{*}$ are linear in $\pi_{1} $ and $\pi_{2} $ .  
 
 We will show now that just as we may construct the usual  scalar electrodynamics by taking a scalar doublet, both this scalars ($\pi_{1} $ and $\pi_{2} $) satisfy identical dynamics  but  no charge.  A charged scalar doublet now allows one to map this doublet in the complex space.

 In our usual representation by complex scalar doublet we chose   the dynamics of the system   by adding the individual Lagrangian as
\begin{equation}
\mathcal{L } = \mathcal{L_1 }+ {\mathcal{L_2 }}
\end{equation} 
 Note however  it will be wrong that the above prescription is the only choice .

  The new scalar electrodynamics will be built on combination  of two elements from the space of the Galileons. How should the elements  combine to meet the very stringent symmetry requirements for the fracton is definitely an intriguing question.
  
  We start with enlisting the symmetries that the new action must posses,

 \begin{enumerate}
 
\item[a.]
 The Lagrangian must be real.

\item[b.]
It must be Poincare invariant.

\item[c.]
It must be symmetric under global phase rotations i.e. we must have  scalar Lagrangian with respect to the transformations $\phi \to e^{i\Lambda } \phi$

\item[d.]
It will have  shift symmetry with the parameters extended to complex values.

\item[e.]
It must be a higher order theory with the equation of motion must be second order in time derivative of $\phi$ and $\phi^{*}$.

\item[f.]
It must be symmetric under the interchange of $\phi$ and $\phi^{*}$ (charge conjugation symmetry).
  \end{enumerate}
  
 We start from a general form   
  \begin{eqnarray}
  {\mathcal{L}}  =  {\mathcal{L}}  (\phi,\phi^{*},\partial_{\mu}\phi,\partial_{\mu}\phi^{*},\partial_{\mu}\partial^{\nu}\phi,\partial_{\mu}\partial^{\nu}\phi^{*})
  \end{eqnarray}
  
\begin{enumerate}  
 \item[1.] 
Due to condition (b) above $\phi$ and $\phi^{*}$  cannot occur explicitly in the Lagrangian otherwise the translation in-variance (which is the inhomogeneous part of the Poincare in-variance) will be lost.

\item[2.]
Again ${\mathcal{L}}$ must be invariant  under Lorentz transformation (which is the homogeneous part of the Poincare in-variance). So all the indices must be properly contracted . 
 
\item[3.] 
 To implement phase rotation symmetry as par (c) $\phi$ and $\phi^{*}$ term must occur in pair. 
 
\end{enumerate}

 So we are left with condition  (d) and (e) . To tackle this point we look at the list of Galileon Lagrangian given above. The structure of these Galileons indicate the process of implementation of the shift symmetry . Read with the condition (c) above  implies that the fourth Galileon Lagrangian ($L_{4}$) is only relevant in our construction because $L_{3}$ and $L_{5}$ contain odd number of $\pi$ term. So these two structure fail to satisfy our recruitment that $\phi$ and $\phi^{*}$ term must occur in pair .

All these condition keep in mind and from the idea of fourth Galileon Lagrangian  we proposed the  Lagrangian as , 

\begin{eqnarray}
 {\mathcal{L}} &=& \bigg[ 2\bigg(\partial_{\mu}\phi\partial^{\mu}\partial_{\nu}\phi\partial^{\nu}\partial_{\rho}\phi^{*}\partial^{\rho}\phi^{*}  - \Box\phi\partial_{\nu}\phi\partial^{\rho}\phi^{*} \partial^{\nu}\partial_{\rho}\phi^{*} \bigg) + 2 \bigg(\partial_{\mu}\phi\partial^{\mu}\partial_{\nu}\phi^{*}\partial^{\nu}\partial_{\rho}\phi\partial^{\rho}\phi^{*}  - \nonumber\\          &  &\qquad \qquad   \Box\phi^{*}\partial_{\nu}\phi^{*}\partial^{\rho}\phi \partial^{\nu}\partial_{\rho}\phi \bigg)+\bigg(\partial_{\mu}\phi^{*}\partial^{\mu}\partial_{\nu}\phi\partial^{\nu}\partial_{\rho}\phi\partial^{\rho}\phi^{*}  - \Box\phi\partial_{\nu}\phi^{*}\partial^{\rho}\phi^{*} \partial^{\nu}\partial_{\rho}\phi \bigg) \nonumber\\          &  &\qquad \qquad  + \bigg(\partial_{\mu}\phi\partial^{\mu}\partial_{\nu}\phi^{*}\partial^{\nu}\partial_{\rho}\phi^{*}\partial^{\rho}\phi  - \Box\phi^{*}\partial_{\nu}\phi\partial^{\rho}\phi \partial^{\nu}\partial_{\rho}\phi^{*} \bigg)\bigg] \label{1}
\end{eqnarray}

See that only mimicking the form of $L_{4}$ (\ref {011}) does not work because our Lagrangian must have the symmetry under  the interchange of $\phi$ and $\phi^{*}$. A little reflection shows that this symmetry may be gain by replicating the form of $L_{4}$ in the indicated manner. Thus the conditions (a) to (f) uniquely define the fracton Lagrangian (\ref{1}). We can thus say that this construction satisfy both co-variance and gauge in-variances. This is definitely hitherto novel and is marked with the universal generalization that make physics so interesting. 

\section{Justification of the proposed Lagrangian}

So we have actually constructed a space-time invariant and gauge invariant action for fracton matter. Our demand of the space-time and gauge invariant Lagrangian   is so brilliant and gratifying that  it calls for a reverse verification that the Lagrangian (\ref{1}) really manifest the symmetry   condition (a) to (f).  
Thus, for instance the shift symmetry of the model is claimed to be 
satisfied by a complex extension of the well known shift symmetry of theories investigated by many stalwarts \cite{sei}\cite{nic}\cite{and}\cite{BR0}\cite{hh}. It is natural for the sceptic to think that the claims are vacuous. So we have verified these issues from different points of view. 

So the first task will be to show by explicit calculations from our model the consequences of all the symmetries. On the top of that we have to show that the equations of motion do not contain any higher derivative term. For the clarity of presentation we will again divide our results in subsections.

\subsection{Equation of motion  }

The specific Lagrangian (\ref{1}) which is a function  $\phi$ and  $\phi^{*}$, of the fields and there derivatives upto the second order . Note that one will expect
that the equations of motion would contain time derivatives of the fields of third order. But the Galileon model is so constructed that such terms will cancel out. The equations of motion are, computed as in the following. The action is, 
\begin{eqnarray} 
 S= \int {\mathcal{L}}  (\phi,\phi^{*},\partial_{\mu}\phi,\partial_{\mu}\phi^{*},\partial_{\mu}\partial^{\nu}\phi,\partial_{\mu}\partial^{\nu}\phi^{*}) d^{4}x \nonumber
\end{eqnarray}

 As usual, let there be a variation $\delta\phi$ which vanishes at the initial and final instants as well as the spatial boundaries (which is usually at infinite). Hence the equation of motion is,

\begin{eqnarray}
 \frac{\delta S}{\delta \phi}= \bigg[\frac{\partial{\mathcal{L}}}{\partial\phi} -\partial_{\mu}\bigg(\frac{\partial{\mathcal{L}}}{\partial(\partial_{\mu}\phi)}\bigg)+\partial_{\mu}\partial^{\nu}\bigg(\frac{\partial{\mathcal{L}}}{\partial(\partial_{\mu}\partial^{\nu}\phi)}\bigg)\bigg] =0\label{4}
\end{eqnarray}

Using the  Lagrangian ({\ref{1}}) , we get the equation of motion of our model as
\begin{eqnarray}
\Box{\phi}\bigg(\partial^{\nu}\partial_\rho \phi^{*} \partial_{\nu}\partial^{\rho} \phi^* - \Box{\phi^{*}} \Box{\phi^{*}}\bigg)+ 2\partial_{\mu}\partial^{\nu}\phi \bigg(\partial^{\mu}\partial_\nu \phi^{*} \Box{\phi^{*}} - \partial^{\mu}\partial_{\rho}\phi^{*}\partial_{\nu}\partial^{\rho}\phi^{*}\bigg)  =0
\label{eom}
\end{eqnarray} 

 Similarly the equation of motion for $\phi^{*}$ may be obtained.  If we substitute $\pi_{2}=0$   and  $\pi_{1}=\pi$ in the equation (\ref{eom}) , then we get
  \begin{eqnarray}
\Box{\pi}\bigg(\partial^{\nu}\partial_\rho \pi \partial_{\nu}\partial^{\rho} \pi - \Box{\pi} \Box{\pi}\bigg)+ 2\partial_{\mu}\partial^{\nu}\pi \bigg(\partial^{\mu}\partial_\nu \pi \Box{\pi} - \partial^{\mu}\partial_{\rho}\pi\partial_{\nu}\partial^{\rho}\pi\bigg)  =0
\end{eqnarray}

 This is a remarkable thing that in this limit our equation of motion reduces in a form which is well known in the Galileon physics \cite{nic}. This can be made more explicit by rearranging terms , we get  
\begin{eqnarray}
(\Box{\pi})^{3}- 3 \Box{\pi} (\partial_{\mu}\partial^{\nu}\pi)^{2} +2 (\partial_{\mu}\partial^{\nu}\pi)^{3} =0
\end{eqnarray} 
  This is the same equation which we see in \cite{nic}\footnote{ See page 9 , equation (42) of \cite{nic}.}. Such a neat cycle of logic is indeed commendable. We have contracted the fracton field  by a combination of the Galileons. But then the Lagrangian for the fracton was derived uniquely from the symmetries of the fracton phase by standard algorithm . The equation of motion (\ref{eom})  thus obtained is shown to pass the Galileon equation of motion in the appropriate limit.
   Thus the theory proposed here can be considered as the complex continuation of Galileon theory \cite{nic}. As far as we know this is the first occurrence of  such extension of Galileon model in the literature.
 
 \subsection{\bf{ Global phase rotation symmetry }}
 
As stated about our model (\ref{1}) is endowed with various symmetries.     The ${\mathcal{U}}(1)$ symmetry is obvious. If we substitute $\phi \to e^{i\Lambda}\phi $ and $\phi^{*} \to e^{-i\Lambda}\phi^{*}  $ in (\ref{1}) then we see that the Lagrangian is unchanged. Carefully note that this will only be true when the parameter $\Lambda$ is constant (i.e only for global phase rotation). Now under infinitesimal phase rotation $\phi$ and $\phi^{*}$ transform as

\begin{eqnarray}
\phi &\to& \phi \bigg(1 + i\Lambda\bigg) \nonumber\\
\phi^{*} &\to &\phi^{*}  \bigg((1 - i\Lambda\bigg) \label{p}
\end{eqnarray}
 putting these value in (\ref{1}) and after some simple but lengthy calculation we easily verified that  Lagrangian (\ref{1}) is invariant under infinitesimal phase rotation (\ref{p}).
 
\subsection{\bf{ Space time symmetry }}

 Slightly more elaborate is the space time symmetry. Under space time transformation the coordinates, fields and their derivatives  transforms as   

 \begin{eqnarray}
   \delta x_{\mu} &=& \xi_{\mu} \nonumber\\
 \delta \phi &=& -\xi^{\lambda}\partial_{\lambda}\phi \nonumber\\
 \delta \phi^{*} &=& -\xi^{\lambda}\partial_{\lambda}\phi^{*} \nonumber\\
\delta (\partial_{\mu}\phi) &=& -\xi^{\lambda}\partial_{\mu}\partial_{\lambda}\phi + \theta_{\mu}{}^{\lambda}\partial_{\lambda}\phi\nonumber\\
\delta (\partial_{\mu}\phi^{*}) &=& -\xi^{\lambda}\partial_{\mu}\partial_{\lambda}\phi^{*} + \theta_{\mu}{}^{\lambda}\partial_{\lambda}\phi^{*}\nonumber\\
\delta (\partial_{\mu}\partial_{\nu}\phi) &=& -\xi^{\lambda}\partial_{\mu}\partial_{\nu}\partial_{\lambda}\phi + \theta_{\mu}{}^{\lambda}\partial_{\nu}\partial_{\lambda}\phi+\theta_{\nu}{}^{\lambda}\partial_{\mu}\partial_{\lambda}\phi\nonumber\\
\delta (\partial_{\mu}\partial_{\nu}\phi^{*}) &=& -\xi^{\lambda}\partial_{\mu}\partial_{\nu}\partial_{\lambda}\phi^{*} + \theta_{\mu}{}^{\lambda}\partial_{\nu}\partial_{\lambda}\phi^{*}+\theta_{\nu}{}^{\lambda}\partial_{\mu}\partial_{\lambda}\phi^{*} \label{pg}
 \end{eqnarray} 
 
 Here , $\xi^{\lambda}=\epsilon^{\lambda}+\theta^{\lambda}{}_{\rho}x^{\rho}$ are the infinitesimal Poincare transformation parameters. Under the transformation  (\ref{pg}) the total change of Lagrangian (\ref{1}) is 
 
\begin{eqnarray}
\Delta  {\mathcal{L}}= \delta {\mathcal{L}} +\xi^{\lambda} \partial_{\lambda}{\mathcal{L}}+\partial_{\lambda}\xi^{\lambda}{\mathcal{L}}
\end{eqnarray} 
 
 where $\delta {\mathcal{L}}$ is the form variation of ${\mathcal{L}}$ . Using the anti-symmetric property of $\theta^{\lambda}{}_{\rho}$  we can easily show that ,
 
 \begin{eqnarray}
\Delta  {\mathcal{L}}= 0 
\end{eqnarray} 
 
  Thus our Lagrangian (\ref{1}) is symmetric under full Poincare group.
 
\subsection{ Shift symmetry }

Now the most nontrivial question  is whether the theory (\ref{1}) has shift symmetry or not. Here the Galileon scalars $\pi_1$
and $\pi_2$ transform as,

\begin{eqnarray} 
 \pi_{1} \to \pi_{1}+ a+b_{\mu}x^{\mu}\nonumber\\
 \pi_{2} \to \pi_{2}+ a+b_{\mu}x^{\mu} \label{12}
  \end{eqnarray}
  
   Using the defining relations (\ref{98})  we obtain

\begin{eqnarray}
\phi &\to& \bigg(\phi +(1+ i) (a+b_{\mu}x^{\mu})\bigg) \nonumber\\
\phi^{*} &\to &\bigg(\phi^{*} + (1 - i) (a+ b_{\mu}x^{\mu})\bigg)\label{mo}
\end{eqnarray} 

Thus  under shift transformation (\ref{mo}) the  fields and their derivatives  transforms as 

\begin{eqnarray}
  \delta\phi &=& (1+i) (a+ b_{\mu}x^{\mu})   \hskip 1cm
 \delta\phi^{*}=(1-i)(a + b_{\mu}x^{\mu}) \nonumber\\
 \delta(\partial_{\mu}\phi)&=&(1+i)b_{\mu}   \hskip 2cm
 \delta(\partial_{\mu}\phi^{*})=(1-i)b_{\mu} \nonumber\\
 \delta(\Box{\phi})&=& 0  \hskip 3cm
 \delta(\Box{\phi^{*}})= 0 \label{s}
 \end{eqnarray}

 The variation of (\ref{1}) due to shift transformation is given by
 
  \begin{eqnarray}
 {\delta\mathcal{L}}&=&\partial^{\mu}\bigg[\{b_{\nu}\partial_{\mu}\partial^{\nu}\phi \partial^{\rho}\phi^{*} - b_{\mu} \Box{\phi} \partial^{\rho}\phi^{*}\}\{(1+i)\partial_{\rho}\phi^{*}+2(1-i)\partial_{\rho}\phi\} \nonumber\\          &  &\qquad \qquad +\{ b_{\nu}\partial_{\mu}\partial^{\nu}\phi^{*} \partial^{\rho}\phi-b_{\mu} \Box{\phi^{*}} \partial^{\rho}\phi\} \{2(1+i)\partial_{\rho}\phi^{*}+      (1-i)\partial_{\rho}\phi\} \bigg] \label{sh}
\end{eqnarray} 

 We observed that the shift invariance of Lagrangian (\ref{1}) is at quasi level because ${\delta\mathcal{L}}$  changes by a partial derivative.

\subsection{Other symmetries}

Right from the beginning of our construction  we have worked with variables that have a well defined transformations under the Poincare group and our Lagrangian is fully contracted. Hence the space time symmetries are all satisfied.

  Of the discrete symmetries, one may be interested to know whether the  charge conjugation symmetry which exist in electrodynamics  is satisfied by our model (\ref{1}). The answer to it will be obtained simply by looking at the Lagrangian.

 We are now satisfied about the  symmetries of the model which are seen to be identical to the phenomenology of fracton . So the achievement of our action  is at par with  the  existing fracton gauge theory \cite{pret1}\cite{nand}. But the possibilities of our model exceeds any existing theory because it is both co-variant and gauge invariant. In this paper we will cite two examples - namely the electromagnetic interaction and gravitational interaction of the fractons. This analysis will show the power and efficacy of the Lagrangian (\ref{1}) in fracton phases. 
 
Before we discuss the applications it will be appropriate to find the conservation laws quantitatively. So in the next section  we will present these calculations from our model  (\ref{1}).

\section{Conservation Laws from different symmetries}

 The fact that we have a covariant as well as gauge invariant theory of fracton phase is itself a great news. We will give an elaborate analysis of the symmetries of our model so that one can understand 
 that why it is obvious that our model may have two independent gauge symmetries. Also this will explain these two symmetries as natural outcome of our model. Needless to say this achievement is  provided the strongest reason for considering our works .

  We have seen that Galileon charged matter  theory  has two   gauge symmetries, the first of these is the well known phase rotation symmetry and the other one is analytic continuation of the shift symmetry.   We must see what conservation laws are indicated by it. A useful  procedure is to apply Noether's theorems. Due to  higher derivatives in the Lagrangian the Noether's formula will be appropriately modified. So  the conservation law is better to be derived from the functional derivative with respect to the parameters of transformation. First let us derive a general formula.  This can then be specialized to different symmetry transformations of (\ref{1}).

          If $\delta\phi$ be the change in $\phi$ then the action  changes by,
\begin{eqnarray}
  \delta S &= & \int\bigg[\frac{\partial{\mathcal{L}}}{\partial\phi}\delta \phi+\frac{\partial{\mathcal{L}}}{\partial\phi^{*}}\delta \phi^{*} +\frac{\partial{\mathcal{L}}}{\partial(\partial_{\mu}\phi)}\delta(\partial_{\mu} \phi)+\frac{\partial{\mathcal{L}}}{\partial(\partial_{\mu}\phi^{*})}\delta(\partial_{\mu} \phi^{*})     \nonumber\\          &  &\qquad \qquad   +\frac{\partial{\mathcal{L}}}{\partial(\partial_{\mu}\partial^{\nu}\phi)}\delta(\partial_{\mu}\partial^{\nu} \phi)+\frac{\partial{\mathcal{L}}}{\partial(\partial_{\mu}\partial^{\nu}\phi^{*})}\delta(\partial_{\mu}\partial^{\nu} \phi^{*})\bigg]d^{4}x \nonumber
   \label {ma}
\end{eqnarray}
Note that $\delta\phi$ not necessarily vanish on the boundary. Using equations of motion we obtain

\begin{eqnarray}
\delta S &=& \int\partial_{\sigma}\bigg[ \frac{\partial{\mathcal{L}}}{\partial(\partial_{\sigma}\phi)}\delta\phi  +\frac{\partial{\mathcal{L}}}{\partial(\partial_{\sigma}\phi^{*})}\delta\phi^{*}  -\partial^{\lambda}\bigg(\frac{\partial{\mathcal{L}}}{\partial(\partial_{\sigma}\partial^{\lambda}\phi)}\bigg)\delta \phi  \nonumber\\          &  &\qquad \qquad -   \partial^{\lambda}\bigg(\frac{\partial{\mathcal{L}}}{\partial(\partial_{\sigma}\partial^{\lambda}\phi^{*})}\bigg)\delta \phi^{*} \bigg] d^{4}x \label{211}
\end{eqnarray}

Equation (\ref{211}) is the master formula for calculating the conserved currents and charges. Now, we will specialize for different symmetries. Start with the shift symmetry . For shift transformation (\ref{mo}) we see , $\delta\phi= (1+ i) (a+ b_{\mu}x^{\mu})$ , $\delta\phi^{*}=(1-i)(a + b_{\mu}x^{\mu})$, substituting this variations of $\phi$ and $ \phi^*$ in (\ref{211}) and after some algebra
we obtain expressions 

\begin{eqnarray}
{\theta}^{\sigma} &=& 3(1+i)(a+ b_{\rho}x^{\rho})\bigg(\partial^{\sigma}\partial_{\mu}\phi\{\partial^{\nu}\phi^{*}\partial_{\nu}\partial^{\mu}\phi^{*}-\partial^{\mu}\phi^{*}\Box{\phi^{*}}\}+\partial^{\sigma}\partial_{\mu}\phi^{*}\{\partial^{\nu}\phi^{*}\partial_{\nu}\partial^{\mu}\phi-\nonumber\\          &  &\qquad \qquad  \partial^{\mu}\phi^{*}\Box{\phi}\} + \partial^{\sigma}\phi^{*} \{ \Box{\phi^{*}}\Box{\phi} - \partial^{\nu}\partial_{\mu}\phi \partial^{\mu}\partial_{\nu}\phi^{*}\}\bigg)+ c.c.  \label{13}
\end{eqnarray}
where $c.c.$ denotes complex conjugation. Thus the current corresponding to $b_\mu$
is given by

\begin{eqnarray}
{\theta^{\sigma\rho}}_{b}&=& x^{\rho}\bigg[3(1+i)\bigg(\partial^{\sigma}\partial_{\mu}\phi\{\partial^{\nu}\phi^{*}\partial_{\nu}\partial^{\mu}\phi^{*}-\partial^{\mu}\phi^{*}\Box{\phi^{*}}\}+\partial^{\sigma}\partial_{\mu}\phi^{*}\{\partial^{\nu}\phi^{*}\partial_{\nu}\partial^{\mu}\phi-\nonumber\\          &  &\qquad \qquad  \partial^{\mu}\phi^{*}\Box{\phi}\} + \partial^{\sigma}\phi^{*} \{ \Box{\phi^{*}}\Box{\phi} - \partial^{\nu}\partial_{\mu}\phi \partial^{\mu}\partial_{\nu}\phi^{*}\}\bigg)+ c.c.\bigg]\label{17}
\end{eqnarray}
Note that the current is a two indexed quantity . So the corresponding charge is 

\begin{eqnarray}
{J^k}_b &=&\int {\theta^{0 k}}_{b}d^3x = \int x^{k}\bigg[3(1+i)\bigg(\partial_{\mu}\dot{\phi}\{\partial^{\nu}\phi^{*}\partial_{\nu}\partial^{\mu}\phi^{*}-\partial^{\mu}\phi^{*}\Box{\phi^{*}}\}+\partial_{\mu}\dot{\phi^{*}}\{\partial^{\nu}\phi^{*}\partial_{\nu}\partial^{\mu}\phi-\nonumber\\          &  &\qquad \qquad  \partial^{\mu}\phi^{*}\Box{\phi}\} + \dot{\phi^{*}} \{ \Box{\phi^{*}}\Box{\phi} - \partial^{\nu}\partial_{\mu}\phi \partial^{\mu}\partial_{\nu}\phi^{*}\}\bigg)+ c.c.\bigg]d^3x
\label{171}
\end{eqnarray}

Again the conserved current corresponding to $a$ is given by

\begin{eqnarray}
{\theta^{\sigma}}_{a} &=& \bigg[3(1+i)\bigg(\partial^{\sigma}\partial_{\mu}\phi\{\partial^{\nu}\phi^{*}\partial_{\nu}\partial^{\mu}\phi^{*}-\partial^{\mu}\phi^{*}\Box{\phi^{*}}\}+\partial^{\sigma}\partial_{\mu}\phi^{*}\{\partial^{\nu}\phi^{*}\partial_{\nu}\partial^{\mu}\phi-\nonumber\\          &  &\qquad \qquad  \partial^{\mu}\phi^{*}\Box{\phi}\} + \partial^{\sigma}\phi^{*} \{ \Box{\phi^{*}}\Box{\phi} - \partial^{\nu}\partial_{\mu}\phi \partial^{\mu}\partial_{\nu}\phi^{*}\}\bigg)+ c.c.\bigg]\label{18}
\end{eqnarray}

So the corresponding charge is 

\begin{eqnarray}
\int {\theta^{0}}_{a}d^3x &=& \int  \bigg[3(1+i)\bigg(\partial_{\mu}\dot{\phi}\{\partial^{\nu}\phi^{*}\partial_{\nu}\partial^{\mu}\phi^{*}-\partial^{\mu}\phi^{*}\Box{\phi^{*}}\}+\partial_{\mu}\dot{\phi^{*}}\{\partial^{\nu}\phi^{*}\partial_{\nu}\partial^{\mu}\phi-\nonumber\\          &  &\qquad \qquad  \partial^{\mu}\phi^{*}\Box{\phi}\} + \dot{\phi^{*}} \{ \Box{\phi^{*}}\Box{\phi} - \partial^{\nu}\partial_{\mu}\phi \partial^{\mu}\partial_{\nu}\phi^{*}\}\bigg)+ c.c.\bigg]  d^3x
\label{172}
\end{eqnarray}

Now for infinitesimal phase rotation (\ref{p})  , $\delta\phi=  i\Lambda\phi$ and $\delta\phi^{*}=-i\Lambda\phi^{*}$, substituting this variations of $\phi$ and $ \phi^*$ in (\ref{211}) and after some algebra
we obtain expressions for the conserved current corresponding to phase rotation symmetry which is given by   ,

\begin{eqnarray}
{\theta^{\sigma}}_{\Lambda} &=& \bigg[3 i \phi \bigg(\partial^{\sigma}\partial_{\mu}\phi\{\partial^{\nu}\phi^{*}\partial_{\nu}\partial^{\mu}\phi^{*}-\partial^{\mu}\phi^{*}\Box{\phi^{*}}\}+\partial^{\sigma}\partial_{\mu}\phi^{*}\{\partial^{\nu}\phi^{*}\partial_{\nu}\partial^{\mu}\phi-\nonumber\\          &  &\qquad \qquad  \partial^{\mu}\phi^{*}\Box{\phi}\} + \partial^{\sigma}\phi^{*} \{ \Box{\phi^{*}}\Box{\phi} - \partial^{\nu}\partial_{\mu}\phi \partial^{\mu}\partial_{\nu}\phi^{*}\}\bigg)+ c.c.\bigg] \label{1800}
\end{eqnarray}

So the corresponding charge is 

\begin{eqnarray}
\int {\theta^{0}}_{\Lambda}d^3x &=&  \int   \bigg[3 i \phi \bigg(\partial_{\mu}\dot{\phi}\{\partial^{\nu}\phi^{*}\partial_{\nu}\partial^{\mu}\phi^{*}-\partial^{\mu}\phi^{*}\Box{\phi^{*}}\}+\partial_{\mu}\dot{\phi^{*}}\{\partial^{\nu}\phi^{*}\partial_{\nu}\partial^{\mu}\phi-\nonumber\\          &  &\qquad \qquad  \partial^{\mu}\phi^{*}\Box{\phi}\} + \dot{\phi^{*}} \{ \Box{\phi^{*}}\Box{\phi} - \partial^{\nu}\partial_{\mu}\phi \partial^{\mu}\partial_{\nu}\phi^{*}\}\bigg)+ c.c.\bigg] d^3x
\label{188}
\end{eqnarray}

An important point to be noted that when the symmetries are localized , there is no difference in the effect of phase rotation and the `translation' part of shift symmetry \footnote{Similar correspondence was reported earlier \cite{and}.}. We see that things may be arranged so that corresponding to a phase rotation there is a shift translation   . Not only that the quantitative relation between the two parameters can easily be taken out as 
\begin{eqnarray}
(1+i) a({\bf{x}}, t) &=& i \Lambda({\bf{x}}, t) \phi({\bf{x}}, t) \nonumber \\
(1-i) a({\bf{x}}, t) &=& -i \Lambda({\bf{x}}, t) \phi^{*}({\bf{x}}, t)\label{201}
\end{eqnarray}

If we consider $A=(1+i) a$ ,$A^{*}=(1-i) a$, $\lambda=i \Lambda$ , $\lambda^{*}=-i \Lambda$ , $B_{\mu}=(1+i) b_{\mu}$ and $B_{\mu}^{*}=(1-i) b_{\mu}$ then from (\ref{201}) we can write ,

\begin{eqnarray}
A({\bf{x}}, t) &=&  \lambda({\bf{x}}, t)  \phi({\bf{x}}, t) \nonumber \\
A^{*}({\bf{x}}, t) &=& \lambda^{*}  ({\bf{x}}, t) \phi^{*}  ({\bf{x}}, t)\label{202}
\end{eqnarray}

\hskip .5cm Now  the  corresponding conserved current for the phase rotation parameter ($\Lambda$) (\ref{1800}) can be written as ,
\begin{eqnarray}
\theta^{\sigma}&=& \frac{\delta S}{\delta \lambda}+\frac{\delta S}{\delta \lambda^{*}}=  \bigg[3  \phi \bigg(\partial^{\sigma}\partial_{\mu}\phi\{\partial^{\nu}\phi^{*}\partial_{\nu}\partial^{\mu}\phi^{*}-\partial^{\mu}\phi^{*}\Box{\phi^{*}}\}+\partial^{\sigma}\partial_{\mu}\phi^{*}\{\partial^{\nu}\phi^{*}\partial_{\nu}\partial^{\mu}\phi-\nonumber\\          &  &\qquad \qquad  \partial^{\mu}\phi^{*}\Box{\phi}\} + \partial^{\sigma}\phi^{*} \{ \Box{\phi^{*}}\Box{\phi} - \partial^{\nu}\partial_{\mu}\phi \partial^{\mu}\partial_{\nu}\phi^{*}\}\bigg)+ c.c.\bigg] \label{183}
\end{eqnarray}

The  charge density is  computed by putting $\sigma = 0$. 
So electric charge is
\begin{eqnarray}
Q &=& \bigg[3 \phi \bigg(\partial_{\mu}\dot{\phi}\{\partial^{\nu}\phi^{*}\partial_{\nu}\partial^{\mu}\phi^{*}-\partial^{\mu}\phi^{*}\Box{\phi^{*}}\}+\partial_{\mu}\dot{\phi^{*}}\{\partial^{\nu}\phi^{*}\partial_{\nu}\partial^{\mu}\phi-\nonumber\\          &  &\qquad \qquad  \partial^{\mu}\phi^{*}\Box{\phi}\} + \dot{\phi^{*}} \{ \Box{\phi^{*}}\Box{\phi} - \partial^{\nu}\partial_{\mu}\phi \partial^{\mu}\partial_{\nu}\phi^{*}\}\bigg)+ c.c.\bigg] d^{3}x \label{188a}
\end{eqnarray}

Naturally this is the charge of the fracton.
This is quite like the usual theory and does not depend on the choice of the Galileons as the basic units.

\hskip .5cm  Similarly for the translation part of shift ($a$) the  conserved current (\ref{18}) can be written as ,

\begin{eqnarray}
{\theta^{\prime\sigma}} &=& \frac{\delta S}{\delta A} + \frac{\delta S}{\delta A^{*}} = \bigg[3\bigg(\partial^{\sigma}\partial_{\mu}\phi\{\partial^{\nu}\phi^{*}\partial_{\nu}\partial^{\mu}\phi^{*}-\partial^{\mu}\phi^{*}\Box{\phi^{*}}\}+\partial^{\sigma}\partial_{\mu}\phi^{*}\{\partial^{\nu}\phi^{*}\partial_{\nu}\partial^{\mu}\phi-\nonumber\\          &  &\qquad \qquad  \partial^{\mu}\phi^{*}\Box{\phi}\} + \partial^{\sigma}\phi^{*} \{ \Box{\phi^{*}}\Box{\phi} - \partial^{\nu}\partial_{\mu}\phi \partial^{\mu}\partial_{\nu}\phi^{*}\}\bigg)+ c.c.\bigg]\label{018}
\end{eqnarray}
  
So the corresponding charge is

\begin{eqnarray}
Q^{\prime} & = & \int  \bigg[3\bigg(\partial_{\mu}\dot{\phi}\{\partial^{\nu}\phi^{*}\partial_{\nu}\partial^{\mu}\phi^{*}-\partial^{\mu}\phi^{*}\Box{\phi^{*}}\}+\partial_{\mu}\dot{\phi^{*}}\{\partial^{\nu}\phi^{*}\partial_{\nu}\partial^{\mu}\phi-\nonumber\\          &  &\qquad \qquad  \partial^{\mu}\phi^{*}\Box{\phi}\} + \dot{\phi^{*}} \{ \Box{\phi^{*}}\Box{\phi} - \partial^{\nu}\partial_{\mu}\phi \partial^{\mu}\partial_{\nu}\phi^{*}\}\bigg)+ c.c.\bigg]  d^3x
\label{0172}
\end{eqnarray}
obtained by putting $\sigma $ = zero .

  We have argued that the symmetry corresponding to  the phase rotation is equivalent to the symmetry corresponding to the shift translation. This equivalence becomes an equality if we note that, on account of (\ref{202}),
 
 \begin{eqnarray}
 \frac{\delta S}{\delta A} &=& \frac{\delta S}{\delta \lambda}\frac{\delta \lambda}{\delta A} = \frac{1}{\phi}\frac{\delta S}{\delta\lambda} \nonumber\\
 \frac{\delta S}{\delta A^{*}} &=& \frac{\delta S}{\delta \lambda^{*}}\frac{\delta \lambda^{*}}{\delta A^{*}} = \frac{1}{\phi^{*}}\frac{\delta S}{\delta\lambda^{*}}
 \end{eqnarray}

  thus  $Q$ and $Q^\prime$ are equal. Also from (\ref{171})  the conserved charge belonging  to shift parameter ($b_{\mu}$) is given by ,
 
\begin{eqnarray}
J^k &=& \int x^{k}\bigg[3 \bigg(\partial_{\mu}\dot{\phi}\{\partial^{\nu}\phi^{*}\partial_{\nu}\partial^{\mu}\phi^{*}-\partial^{\mu}\phi^{*}\Box{\phi^{*}}\}+\partial_{\mu}\dot{\phi^{*}}\{\partial^{\nu}\phi^{*}\partial_{\nu}\partial^{\mu}\phi-\nonumber\\          &  &\qquad \qquad  \partial^{\mu}\phi^{*}\Box{\phi}\} + \dot{\phi^{*}} \{ \Box{\phi^{*}}\Box{\phi} - \partial^{\nu}\partial_{\mu}\phi \partial^{\mu}\partial_{\nu}\phi^{*}\}\bigg)+ c.c.\bigg]d^3x\label{0171}
\end{eqnarray}

 The implication of (\ref{0171})
is indeed gratifying. From  (\ref{0172})
 we can write
 \begin{eqnarray}
 Q=Q^{\prime} = \int \rho({\bf{x }},t) d^3{\bf{x}}
 \end{eqnarray}
 where,
\begin{eqnarray}
\rho({\bf{x}}, t)
  & =&\bigg[ 3\bigg(\partial_{\mu}\dot{\phi}\{\partial^{\nu}\phi^{*}\partial_{\nu}\partial^{\mu}\phi^{*}-\partial^{\mu}\phi^{*}\Box{\phi^{*}}\}+\partial_{\mu}\dot{\phi^{*}}\{\partial^{\nu}\phi^{*}\partial_{\nu}\partial^{\mu}\phi-\nonumber\\          &  &\qquad \qquad  \partial^{\mu}\phi^{*}\Box{\phi}\} + \dot{\phi^{*}} \{ \Box{\phi^{*}}\Box{\phi} - \partial^{\nu}\partial_{\mu}\phi \partial^{\mu}\partial_{\nu}\phi^{*}\}\bigg)+ c.c.\bigg]
   \end{eqnarray}

its the charge density. From (\ref{0171}) we can now write the second conserved charge as 
 \begin{eqnarray}
 J^k = \int x^k \rho({\bf{x }},t) d^3{\bf{x}}
 \end{eqnarray}
 Evidently, the second conserved charge is the dipole moment of the system. As promised in the beginning, our choice of fracton matter action along with the application of the gauge principle  derives systematically the conserved quantities.

 \section{  Localization of phase rotation symmetry and complex shift symmetry }
 
 We will employ the celebrated gauge principle to localize the different symmetry of our model.  It will be convenient to introduce the concept of ``gauge principle" in the familiar context of Maxwell's electromagnetic theory . We start with a free theory namely the zero mass Klein-Gordon Lagrangian 
 \begin{eqnarray}
 {\mathcal{L}_{kg}} = \frac{1}{2}\partial_\mu\phi \partial^\mu\phi^{*} \label{99}
\end{eqnarray}  
  This is symmetric under the global gauge transformation $\phi \to \phi e^{i\Lambda} $, where $\Lambda$ is a constant. The construction of the un-gauged theory is most important for localization process . Note that  here  the particular transformations of $\phi$ and $\partial_{\mu}\phi$ are such  that the Lagrangian (\ref{99}) is  invariant. Indeed, under the infinitesimal phase rotation,
  \begin{eqnarray}
  \delta \phi = i\Lambda \phi    \hskip .2cm \delta{(\partial_
   \mu
  \phi)} = -i\Lambda \partial_\mu\phi \nonumber\\
  \delta \phi^* = i\Lambda \phi^*  \hskip .2cm \delta{(\partial_
   \mu
  \phi^*)} = -i\Lambda \partial_\mu\phi^*
  \label{hybrid}
  \end{eqnarray}
    
 and this is the way the fields and their derivatives should transform so as to ensure invariance.
    
    When the phase rotation symmetries are localized
    the 
     parameters of the  transformation are no longer constant but arbitrary functions of space time. So we consider  $\Lambda$ to be a function of $x$ and $t$. Then  the transformation of the field $\delta\phi$ retains it form but derivatives of it deviates from 
  how they would do earlier. To compensate this change we have to replace the partial derivatives by 
 \begin{eqnarray}
  D_{\mu} {\phi}= 
  (\partial_{\mu}+ iA_\mu)\phi
\end{eqnarray}  
 
 where $A_\mu$ is called a gauge field . Its transformation under the local gauge transformation is assumed so that $\nabla_{\mu}$ transform just as $\partial_{\mu}$ transform in un - gauged theory (\ref{hybrid}). Invariance is thus retained . This is the gauge principle which we see is instrumental to obtain a Lagrangian invariant under local transformation. It is observed that the commutator of $\nabla_{\mu}$ and $\nabla_{\nu}$ satisfy 
 
 \begin{eqnarray}
 [ D_{\mu},D_{\nu} ]= F_{\mu\nu}=\partial_{\mu}A_{\nu} - \partial_{\nu}A_{\mu} \label{101}
\end{eqnarray} 
 and the dynamics of the field $A_{\mu}$ is completely given by $F_{\mu\nu}$. Also it is anti-symmetric over the interchange of   $\mu $ and $\nu$. It is then straight forward to identify the interaction generated by $ A_\mu $ as the electromagnetic force . This is nothing new but a well known fact. The gauge principle is thus a very useful tool in model building. In the following it will be seen that the gauge principle has similar ubiquity in constructing the fracton gauge theory \footnote{From what has been stated in the above, one should not get the impression that the whole electromagnetic theory can be thus constructed. It is rather the opposite. It took more than 200 years of experimenting and theorizing to know $U(1)$ is the appropriate symmetry.}.
 
\hskip .5cm In order to understand the dynamics of the fracton phase of matter in the  electromagnetic field, we have 
 to put the system in space and watch its development in time. Now, if the symmetry elements are uniform i.e the symmetries are global then there is no interaction. However, if the elements change from one point to another new gauge fields will appear as we have seen in the above example. Of the symmetries derived in  section 2, the phase rotation symmetry and  shift  symmetry are manifested in the flat background where the theory is formulated . So confining ourselves in flat background the two internal symmetries should be simultaneously localized because both  transformation takes place in the $\phi$ space. Also at the first glance it appears that  there should be three gauge fields  corresponding to the three parameter ($\Lambda$ ,  $a$, $b_{\mu}$). However both $\Lambda$ and $a$ produce the same effects . Hence we have to introduce only one gauge field corresponding to these two parameter. So the new gauge fields will be two in number which we are denoted by $A_{\mu}$ and $ B_{\mu\nu}$. Thus the co-variant derivatives is given by

\begin{eqnarray}
 D_{\mu} {\phi}= \partial_{\mu}\phi +iA_{\mu}\phi+(1+i)B_{\mu\nu} x^{\nu}\phi
 \label{ko1}
\end{eqnarray}
 Here $ A_\mu $ is a vector under Lorentz transformations and $B_{\mu\nu}$ is second rank tensor .  A little exercise proves that $ B_{\mu\nu}$ may be assumed  symmetric, as is shown in the following .

   Let it be a general second rank tensor. Consider a virtual displacement $ { \delta x^\mu } $ of the system  then from (\ref{ko1}) the contraction 
  
  \begin{eqnarray}
\delta x^{\mu} D_{\mu} {\phi}= \delta x^{\mu}\left(\partial_{\mu}\phi \right ) +i\delta x^{\mu}A_{\mu}\phi+(1+i)\delta x^{\mu} B_{\mu\nu} x^{\nu}\phi
 \label{ko2}
 \end{eqnarray}

 Now any general second rank tensor can be split up in  anti-symmetric and symmetric parts 
 
 \begin{eqnarray}
  B_{\mu\nu}=  \left[B^{(A)}_{\mu\nu} + B^{(S)}_{\mu\nu} \right]  
  \end{eqnarray}

 where 
 \begin{eqnarray}
   B^{(A)}_{\mu\nu} =\frac{1}{2}  \bigg(B_{\mu\nu}- B_{\nu\mu}\bigg)\nonumber\\ 
  B^{(S)}_{\mu\nu} = \frac{1}{2}  \bigg(B_{\mu\nu}+ B_{\nu\mu}\bigg)
 \end{eqnarray}
 
 The left hand side of (\ref{ko2}) is interpreted as the change of the field in the local translation  $ \delta x^\mu $.  One or two steps of algebra show that the anti symmetric part contributes nothing to the factor $\delta x^{\mu} D_{\mu} {\phi}$.  
 So any anti symmetric part of  $ B_{\mu\nu}$ is irrelevant. It implies that  $ B_{\mu\nu}$ may be   assumed  symmetric in (\ref{ko1}) .
 
 \hskip 1.5cm Coming back to the discussion ongoing from the co-variant derivatives (\ref{ko1}), we calculate variations   as  
 
 \begin{eqnarray}
 \delta \phi &=& \bigg(i\Lambda \phi  +(1+i)b_{\mu}x^{\mu}\bigg) \hskip 1cm
 \delta \phi^{*} = \bigg(-i\Lambda \phi^{*}  +(1-i)b_{\mu}x^{\mu}\bigg) \nonumber\\
\delta (\partial_{\mu}\phi) &=& \bigg(i\Lambda \partial_{\mu}\phi  +(1+i)b_{\mu}\bigg) \hskip 1cm
\delta (\partial_{\mu}\phi^{*}) = \bigg(-i\Lambda \partial_{\mu}\phi^{*}  +(1-i)b_{\mu}\bigg) \nonumber \\
\delta (\partial^{\mu}\partial_{\mu}\phi) &=& \bigg(i\Lambda \partial^{\mu}\partial_{\mu}\phi\bigg) \hskip  3 cm
\delta (\partial^{\mu}\partial_{\mu}\phi^{*}) = \bigg(-i\Lambda \partial^{\mu}\partial_{\mu}\phi^{*}\bigg) \label{kolk}
\end{eqnarray}
 
 The above analysis gives us the general methods of  gauging the symmetry and  we have listed the variations of the fields and their derivatives under the {\bf{global}} phase rotation symmetry and  the {\bf {complex shift}} symmetry .

 \hskip 1.5cm  After localization      the transformations of the new gauge fields  $A_{\mu}$ and $ B_{\mu\nu}$  must ensure that the``new derivative ($D_{\mu}\phi$)" will transform in the same way as the usual derivatives (\ref{kolk}) do in the global theory. Thus the variation  of $D_{\mu}\phi$ must be 
 
  \begin{eqnarray}
\delta (D_{\mu}\phi) &=& \bigg(i\Lambda D_{\mu}\phi  +(1+i)b_{\mu}\bigg) \label{ko}
\end{eqnarray}

But from definition of $D_{\mu}\phi$ (\ref{ko1}) we can compute it in terms of the fields independently which should be given by (\ref{ko}), according to gauge principle. In that process by equating the real and imaginary part separately we get two equation involving $\delta A_{\mu}$ and $\delta B_{\mu\nu}$, from which we get,  
\begin{eqnarray}
\phi\delta A_{\mu} &=& -\bigg(\phi\partial_{\mu}\Lambda  + 2 b_{\nu}x^{\nu} A_{\mu} +2 B_{\mu\nu}b_{\alpha}x^{\alpha}x^{\nu}\bigg) \nonumber\\
\phi\delta B_{\mu\nu} &=& \bigg( A_{\mu}b_{\nu}-\partial_{\mu}b_{\nu} \bigg) \label{lo}
\end{eqnarray} 
  This conditions (\ref{lo}) also ensure the co-variance of $D_{\mu}\phi^{*}$.

\hskip 1.5cm Our  complex Galileon model (\ref{1}) also contains second derivative of $\phi$ and $\phi^{*}$. The transformations of the second derivatives (\ref{kolk}), however, do not have any shift symmetry contribution . Consequently,
the second derivative $\partial_\mu\partial^{\mu}\phi$
should be replaced by 
${\bar{D}}_\mu D^{\mu}\phi$ .

Where ${\bar{D}} \phi$ is defined as  
\begin{eqnarray}
{\bar{ D_{\mu}}} {\phi}= \partial_{\mu}\phi +iA_{\mu}\phi
 \label{ko5}
 \end{eqnarray} 
 
 Explicit calculation shows that under complex  shift   and phase rotation transformations  ${\bar{D}}_\mu D^{\mu}\phi$ transform in the same way as $\partial_\mu\partial^{\mu}\phi$ i.e.
\begin{equation}
\delta({\bar{D}}_\mu D^{\mu}\phi)= \bigg(i\Lambda{\bar{D}}_{\mu}D^{\mu}\phi\bigg)
\end{equation}


 In the above we have found out that both the phase
 rotations  and the complex shift symmetries have appeared in global gauge theory. These symmetries 
together define fracton interaction with the applied electromagnetic fields.

\section{ Maxwell's theory in the fracton representation } 

In the last section we have seen that  the localization process applied to the phase rotation symmetry and the shift symmetry    produces two independent gauge fields of which $A_{\mu}$ represent the Maxwell's electrodynamics  and another is  $B_{\mu\nu}$ whose physical content is yet to be relieved . This may be the counter part of the symmetric tensor gauge field in the existing theory.  Here  we derived it in full tonsorial form without any {\it{ad-hoc}} approximation. For instance we have proved that  this tensor must be symmetric (see the text below (\ref{ko1})) whereas in the existing theory it was assumed only.

\hskip .5cm Proceeding with the localization process as above we get the covariant derivative (\ref{ko1}). Now  the commutator of two covariant derivative give the field tensor from which the dynamics can be obtained.   Now  the commutator of $D_{\mu}$ and $D_{\nu}$ are given by
 
 \begin{eqnarray}
 [ D_{\mu},D_{\nu} ]=i\bigg(\partial_{\mu}A_{\nu} - \partial_{\nu}A_{\mu}\bigg)+(1+i)(\partial_{\mu}B_{\nu\rho}-\partial_{\nu}B_{\mu\rho})x^{\rho} + (B_{\mu\nu}- B_{\nu\mu}) \label{100}
\end{eqnarray} 
 As $ B_{\mu\nu}$ is  symmetric tensor  thus  

 \begin{eqnarray}
 [ D_{\mu},D_{\nu} ]=i\bigg(\partial_{\mu}A_{\nu} - \partial_{\nu}A_{\mu}\bigg)+(1+i)(\partial_{\mu}B_{\nu\rho}-\partial_{\nu}B_{\mu\rho})x^{\rho}  \label{1100}
\end{eqnarray} 

To construct the dynamics of the gauge field in our theory we will  introduce   $\mathcal{F_{\mu\nu}}$  where $\mathcal{F_{\mu\nu}}$ is given by 

\begin{eqnarray}
 \mathcal{F_{\mu\nu}}=[ D_{\mu},D_{\nu} ] \label{102}
\end{eqnarray}

Using the equation (\ref{1100}), we get 

\begin{eqnarray}
 \mathcal{F_{\mu\nu}}= F_{\mu\nu} + G_{\mu\nu\rho}x^{\rho} \label{103}
\end{eqnarray} 

Where 
\begin{eqnarray}
 F_{\mu\nu} &=& i(\partial_{\mu}A_{\nu} - \partial_{\nu}A_{\mu})\nonumber \\
 G_{\mu\nu\rho}&=&(1+i)(\partial_{\mu}B_{\nu\rho}-\partial_{\nu}B_{\mu\rho})\label{104}
\end{eqnarray}

 From (\ref{104}) we see that $F_{\mu\nu}$  is anti-symmetric and after some little calculation we can prove that   $ G_{\mu\nu\rho}$ is also totally anti-symmetric. To show this we proceed as below.

 \hskip .25cm If we interchange $\mu$ and $\nu$ then we get 
   \begin{eqnarray}
  G_{\nu\mu\rho}&=&(1+i)(\partial_{\nu}B_{\mu\rho}-\partial_{\mu}B_{\nu\rho})=-G_{\mu\nu\rho}\label{105}
\end{eqnarray}
again  from the definition of $ G_{\nu\mu\rho}$  we get

 \begin{eqnarray}
  G_{\rho\mu\nu}&=&(1+i)(\partial_{\rho}B_{\mu\nu}-\partial_{\mu}B_{\rho\nu})\nonumber \\
   G_{\nu\rho\mu}&=&(1+i)(\partial_{\nu}B_{\rho\mu}-\partial_{\rho}B_{\nu\mu})
\end{eqnarray}
 
 adding these two term we get 
 
  \begin{eqnarray}
  G_{\rho\mu\nu}+G_{\nu\rho\mu}&=&(1+i)\bigg[(\partial_{\rho}B_{\mu\nu}-\partial_{\rho}B_{\nu\mu})-(\partial_{\mu}B_{\rho\nu}-(\partial_{\nu}B_{\rho\mu})\bigg]\nonumber 
\end{eqnarray}
as  $B_{\mu\nu}$ is a symmetric tensor thus we get 

\begin{eqnarray}
  G_{\rho\mu\nu}+G_{\nu\rho\mu}= G_{\nu\mu\rho}
\end{eqnarray}
 
 using (\ref{105})
 
 \begin{eqnarray}
  G_{\rho\mu\nu}+G_{\nu\rho\mu}+ G_{\mu\nu\rho}=0 \label{106}
\end{eqnarray}
 
 
which can be compactly written as 
\begin{equation}
\epsilon^{\lambda\mu\nu\rho}
G_{\mu\nu\rho}  =0
\label{dual}
\end{equation}

 Thus  $G_{\mu\nu\rho}$ must be a anti-symmetric tensor. Now from (\ref{103}) we can say that $\mathcal{F_{\mu\nu}}$ is also a anti-symmetric tensor. We now defined the dual of $\mathcal{F_{\mu\nu}}$, which is given by 
 
 \begin{eqnarray}
  \tilde{\mathcal{F^{\mu\nu}}}  = \epsilon^{\mu\nu\alpha\beta} \mathcal{F_{\alpha\beta}}
\end{eqnarray}

using (\ref{103}) the partial derivative of $\tilde{\mathcal{F^{\mu\nu}}}$  can be written as 

\begin{eqnarray}
 \partial_{\mu} \tilde{\mathcal{F^{\mu\nu}}}  = \epsilon^{\mu\nu\alpha\beta} \bigg[\partial_{\mu}F_{\alpha\beta} + \partial_{\mu}(G_{\alpha\beta\rho}x^{\rho})\bigg]=\epsilon^{\mu\nu\alpha\beta} \bigg[\partial_{\mu}F_{\alpha\beta} + \partial_{\mu}(G_{\alpha\beta\rho})x^{\rho}+G_{\alpha\beta\mu}\bigg]\label{107}
\end{eqnarray}

using the definition of $F_{\alpha\beta}$ and $G_{\alpha\beta\rho}$  and using (\ref{dual}) we can easily show that 
\begin{eqnarray}
 \partial_{\mu} \tilde{\mathcal{F^{\mu\nu}}}  = 0 \label{108}
\end{eqnarray}

Thus in analogy with Maxwell's theory we can define electric field and magnetic field as

\begin{eqnarray}
 \mathcal{E}^{l}&=& \mathcal{F}^{{l 0}}= F^{l 0} + G^{l 0 k}x_{k}=i(\partial^{l}A^{0} - \partial^{0}A^{l})+(1+i)(\partial^{l}B^{0 k}-\partial^{0}B^{l k})x_{k}\nonumber\\
 \mathcal{B}^{l}&=& \tilde{\mathcal{F}^{{l 0}}}= \epsilon^{l 0 m n}\bigg[F_{m n} + G_{m n k}x^{k}\bigg]=\epsilon^{l 0 m n}\bigg[i(\partial_{m}A_{n} - \partial_{n}A_{m})+ \nonumber\\          &  &\qquad \qquad (1+i)(\partial_{m}B_{n k}-\partial_{n}B_{m k})x^{k}\bigg] \label{109}
 \end{eqnarray}

In terms of field tensor $\mathcal{F}^{{\mu\nu}}$ ,   two in-homogeneous Maxwell's equation can be written as 

\begin{eqnarray}
\partial_{\mu} \mathcal{F}^{{\mu\nu}}=  J^{\nu} \label{110}
 \end{eqnarray}
While
the two homogeneous  equations can be written as 

\begin{eqnarray}
 \partial_{\mu} \tilde{\mathcal{F^{\mu\nu}}}  = 0 
\end{eqnarray}

again as $\mathcal{F}^{{\mu\nu}} $ is anti-symmetric, thus  

\begin{eqnarray}
\partial_{\nu} \partial_{\mu} \mathcal{F}^{{\mu\nu}}= 0
 \end{eqnarray}
 thus using (\ref{110}) we can write 
 
\begin{eqnarray}
\partial_{\nu} J^{\nu} =0 \label{111}
 \end{eqnarray}
 
This is the equation of continuity. 
 We can then write all the relevant equations
of electromagnetism in terms of the matrix ${\mathcal{F}}^{\mu\nu}$ as they are expressed in equation (\ref{108}) ,(\ref{110}) .   Hence the theory generated by the tensor gauge field $B_{\mu\nu}$ is nothing but the electromagnetic theory but expressed in a representation different from the usual one. In this representation the complex scalar field introduced by us  determine the matter action. So the concept of generalized Maxwell's theory is a misnomer . Actually it is the electromagnetic interaction in the different, Galileon based representation.  This difference with the existing theory is not to contradict but compliment the existing fracton theory. 

 \section{ Gravitational effect of fractons  }
 There is a conjecture in the existing fracton gauge theory that as  fracton theory is formulated in terms of   a second rank symmetric tensor 
 , it may have a gravitational effect \cite{pret1} \cite{pret2}.  Interestingly they used the same tensor field introduced in connection with electromagnetic interaction of fractons  
 ,as responsible for the emergence of gravity  . This is a singular assumption. Moreover , in combination with the other conclusions it should be looked upon as  a very difficult paradox that whether it suggests  a connection between electromagnetic theory and general theory of relativity or not? It  is a paradox of the existing fracton gauge theory which can never be resolved in the confines of the same theory.

\hskip.5cm All these questions and paradoxes associated with the  fracton - gravity interaction in the existing theory are absent in our formulation, based as it is on an action which has two general principles  co-variance and gauge in-variances. To see how gravitational interaction of the fractons originate in our theory
the following discussion may be helpful. A very quick revision of certain concepts  will be necessary for a proper understanding of our work, specially for those who are not well versed in the gauge theories of gravity .

\hskip.5cm  Einstein, in the formulation of General relativity (GR) considered space-time , the  arena where massive bodies move in a self consistent way. In other words, space-time geometry is considered
as a dynamical object. Massive bodies otherwise free from interaction follow a `straight line' (geodesic) in the space-time. If space time is curved, the geometry is non euclidean.
The curvature of the geometry is obtained from a fourth rank tensor, called the Riemann tensor
 ${R^\gamma} _{\mu\nu \rho} $. In the Gauge theoretic formulation of gravity  Riemann tensor is defined  by the commutator of two covariant derivative \cite{blago}. From the Riemann 
tensor, we get the Richie tensor
\begin{eqnarray}
               R_{\mu\nu} = {R^{\lambda}}_{\mu\lambda\nu} \label{Ri}
\end{eqnarray}
A nonzero Riemann tensor is a guarantee that the space-time is curved space time which is a signature of gravity.  How the matter interacts with the gravitational field is given by the Einstein's equation(in natural units $c=\hbar =1$) ,
\begin{eqnarray}
R_{\mu\nu}-\frac{1}{2}R g_{\mu\nu}=8\pi G T_{\mu\nu}
\label{EEq}
\end{eqnarray}

where the symmetric energy-moment tensor is defined by $T_{\mu\nu}$
 
 \begin{eqnarray}
 T_{\mu\nu} = \frac{1}{\sqrt{-g}}\frac{\delta S}{\delta g^{\mu\nu}} \label{gr0}
\end{eqnarray}

 here $S$ is the action of the matter theory. Often the energy momentum tensor obtained from Noether's theorem agrees with that obtain from (\ref{gr0}). But in many cases they may differ . This difference can again be exploited to obtained significant result \cite{Bp1}. We will come to it in some later communication. 

The method described so far  is the metric approach which is not very suitable for formulating the gauge theory of gravity \cite{blago}. So, the approach to gravity useful for us are however, vielbein based  Poincare gauge theory (PGT) . The PGT formulates gravity   in terms of vielbeins and spin connections. But there are well defined procedures in the literature which connects the 
metric with the vielbein and the affine connection \cite{blago} in terms  of both vielbeins and spin connections. 
The essence of PGT is gauging of symmetries of a Poincare invariant matter theory  \cite{Uti} , \cite{kibble}, \cite {Schima}. From the localization of the shift symmetry of a Poincare invariant theory we get a geometrical manifold (Einstein-Cartan manifold ) which goes to Riemannian geometry under the assumption of symmetric connections \cite{rbp}. This approach is sometimes referred to as the `seventh route to geometrodynamics' \cite{kiefer}.

 \hskip .5cm Now in PGT corresponding to the basic fields $\Sigma^i_{\ \mu}$ and $M^{ij}_{\ \ \mu}$ the Lorentz field strength $R^{ij}_{\ \ \mu\nu}$ and the translation field strength $T_{i\mu\nu}$ are obtained following the usual procedure in gauge theory. The commutator of two covariant derivatives gives \cite{blago} 
\begin{eqnarray}
\left[\nabla_k,\nabla_l\right]\phi = \frac{1}{2} ~\Sigma_k^{\ \mu} \Sigma_l^{\ \nu} ~R^{ij}_{\ \ \mu\nu}\sigma_{ij}\phi - \Sigma_k^{\ \mu}\Sigma_l^{\ \nu}~T^i_{\ \mu\nu}\nabla_i\phi \label{cova}
\end{eqnarray}
These defining equations give the following expressions for the field-strengths
\begin{eqnarray}
\label{pgt}
T^i_{\ \mu\nu} &=& \partial_\mu \Sigma^i_{\ \nu} + M^{i}_{\ \, k\mu} \Sigma^k_{\ \nu} - \partial_\nu \Sigma^i_{\ \mu} - M^{i}_{\ \,k\nu} \Sigma^k_{\ \mu}\nonumber\\
R^{ij}_{\ \ \mu\nu} &=& \partial_\mu M^{ij}_{\ \ \nu} - \partial_\nu M^{ij}_{\ \ \mu} + M^i_{\ k\mu}M^{kj}_{\ \ \nu} - M^i_{\ k\nu}M^{kj}_{\ \ \mu}.
\end{eqnarray}
and also
\begin{eqnarray}
 R^{\rho\sigma}{}_{\mu\nu} = \Sigma_{i}{}^{\rho}\Sigma_{j}{}^{\sigma}  R^{ij}{}_{\mu\nu} \label{rici}
 \end{eqnarray}

So far the theory is in the Minkowski space and has been developed as a gauge theory. From the point of view of geometric interpretation, the Lorentz field strength $R^{ij}_{\ \ \mu\nu}$ and the translation field strength $T_{i\mu\nu}$, correspond to the Riemann tensor and the torsion (see
\cite{blago}). Using these basic structures, gravity can be formulated in the framework of PGT. Now for zero torsion (\ref{cova}) becomes

\begin{eqnarray}
\left[\nabla_k,\nabla_l\right]\phi = \frac{1}{2} ~\Sigma_k^{\ \mu} \Sigma_l^{\ \nu} ~R^{ij}_{\ \ \mu\nu}\sigma_{ij}\phi  \label{cova1}
\end{eqnarray}

But  before localizing the space time symmetry we first show that our Lagrangian (\ref{1}) is invariant under combined Poincare and shift transformations. Now under combined Poincare and shift transformation the fields and their derivative transforms as \cite{rbp}

 \begin{eqnarray}
 \delta \phi &=& -\xi^{\lambda}\partial_{\lambda}\phi + (1+i)b_{\mu}x_{\mu}\nonumber\\
 \delta \phi^{*} &=& -\xi^{\lambda}\partial_{\lambda}\phi^{*} + (1-i)b_{\mu}x_{\mu}\nonumber\\
\delta (\partial_{\mu}\phi) &=& -\xi^{\lambda}\partial_{\mu}\partial_{\lambda}\phi + \theta_{\mu}{}^{\lambda}\partial_{\lambda}\phi+ (1+i)b_{\mu}\nonumber\\
\delta (\partial_{\mu}\phi^{*}) &=& -\xi^{\lambda}\partial_{\mu}\partial_{\lambda}\phi^{*} + \theta_{\mu}{}^{\lambda}\partial_{\lambda}\phi^{*}+ (1-i)b_{\mu}\nonumber\\
\delta (\partial_{\mu}\partial_{\nu}\phi) &=& -\xi^{\lambda}\partial_{\mu}\partial_{\nu}\partial_{\lambda}\phi + \theta_{\mu}{}^{\lambda}\partial_{\nu}\partial_{\lambda}\phi+\theta_{\nu}{}^{\lambda}\partial_{\mu}\partial_{\lambda}\phi\nonumber\\
\delta (\partial_{\mu}\partial_{\nu}\phi^{*}) &=& -\xi^{\lambda}\partial_{\mu}\partial_{\nu}\partial_{\lambda}\phi^{*} + \theta_{\mu}{}^{\lambda}\partial_{\nu}\partial_{\lambda}\phi^{*}+\theta_{\nu}{}^{\lambda}\partial_{\mu}\partial_{\lambda}\phi^{*} \label{pg1}
 \end{eqnarray} 
 
 Here , $\xi^{\lambda}=\epsilon^{\lambda}+\theta^{\lambda}{}_{\rho}x^{\rho}$ are the infinitesimal Poincare transformation parameters. Under the transformation  (\ref{pg1}) the total change of Lagrangian (\ref{1}) is 
 
\begin{eqnarray}
\Delta  {\mathcal{L}}= \delta {\mathcal{L}} +\xi^{\lambda} \partial_{\lambda}{\mathcal{L}}+\partial_{\lambda}\xi^{\lambda}{\mathcal{L}}
\end{eqnarray} 
 
 where $\delta {\mathcal{L}}$ is the form variation of ${\mathcal{L}}$ . Now using the variation (\ref{pg1})  we easily shows that
 
 \begin{eqnarray}
\Delta  {\mathcal{L}} &=& \partial^{\mu}\bigg[\{b_{\nu}\partial_{\mu}\partial^{\nu}\phi \partial^{\rho}\phi^{*} - b_{\mu} \Box{\phi} \partial^{\rho}\phi^{*}\}\{(1+i)\partial_{\rho}\phi^{*}+2(1-i)\partial_{\rho}\phi\} \nonumber\\          &  &\qquad \qquad +\{ b_{\nu}\partial_{\mu}\partial^{\nu}\phi^{*} \partial^{\rho}\phi-b_{\mu} \Box{\phi^{*}} \partial^{\rho}\phi\} \{2(1+i)\partial_{\rho}\phi^{*}+      (1-i)\partial_{\rho}\phi\} \bigg] 
\end{eqnarray}
 
 Thus our Lagrangian (\ref{1}) is quasi-invariant under combined Poincare and shift transformation. 
 
 \hskip .5cm Now we are ready to localized the space-time symmetry  . In order to localized Poincare symmetry we have to replace $\partial_{\mu} $ by

 \begin{eqnarray}
\nabla_{a} \phi= \Sigma_{a}{}^{\mu} D_{\mu} \phi= \Sigma_{a}{}^{\mu} \bigg(\partial_{\mu}\phi+ \frac{1}{2} M_{\mu}{}^{b c}\sigma_{b c}\phi\bigg)
\end{eqnarray}
 
 where $\sigma_{ab}$ is the Lorentz spin matrix  and   $\Sigma_{a}{}^{\mu}$  and $M_{\mu}{}^{bc}$ are the new gauge fields corresponding to translation and Lorentz transformation \footnote{As $\phi$ is a Lorentz scalar , the gauge fields corresponding to the Lorentz rotation do not appear . }. As The Galileon field has an additional shift symmetry , so we require to introduce further gauge field  $P_{\mu\nu}$. The new covariant derivatives are now defined as,
   \begin{eqnarray}
   {\bar{\nabla}}_a\phi = \Sigma_a{}^\mu{\bar{ D}}_\mu \phi
   \label{gf1}
   \end{eqnarray}
   where
   \begin{eqnarray}  
   {\bar{ D}}_\mu \phi = \left(D_\mu\phi + F_\mu\phi\right); \left(F_\mu =(1+i)P_{\mu\nu} x^{\nu} \right)\label{10}
   \end{eqnarray}
  The transformations of the new fields are obtained by demanding that the covariant derivatives (\ref{gf1}) transform as the ordinary ones do in the ungauged theory (\ref{pg1}),

 \begin{equation}
 \delta({\bar{\nabla}}_a\phi)=-\xi^c \partial_c({\bar{\nabla}}_a
 \phi)+{\theta_a}^b{\bar{\nabla}}_b\phi
 +(1+i)b_a
 \end{equation}
 This yields,
 \begin{eqnarray}
 \delta\Sigma_a{}^\mu &=& -\xi^\lambda\partial_\lambda
 \Sigma_a{}^\mu
+\partial_\lambda\xi^\mu
\Sigma_a{}^\lambda + \theta_a
{}^b\Sigma_b{}^\mu\nonumber\\
\delta M_\mu{}^{ab} &=& -\xi^\lambda\partial_\lambda
M_\mu{}^{ab}-\partial_\mu
\theta^{ab}
-\partial_\mu\xi^\lambda M_\mu{}^{ab} + \theta^a{}_c M_\mu{}^{cb} +\theta^b{}_c M_\mu{}^{ac} \nonumber\\
\delta \left(
x^{\nu} P_{\mu\nu}\phi\right) &=& -\xi^\lambda\partial_
\lambda\left(
x^{\nu}P_{\mu\nu}\phi \right) -\partial_\mu
\xi^\lambda \left(
x^{\nu} P_{\lambda\nu}\phi\right) -x^\nu\partial_\mu b_\nu \label{trans}
\end{eqnarray}

 The Galileon symmetries manifested through these relations are compatible with analogous results in \cite{rbp}\cite{GHJT} where the algebra of Galileon generators has been defined.
Also note that 
\begin{eqnarray}
\delta({\bar{D}}_{\mu} \phi) & = &-\xi^{\lambda}\partial_{\lambda}({\bar{D}}_{\mu} \phi)-\partial_{\mu}\xi^{\lambda}{\bar{D}}_{\lambda} \phi+(1+i)b_{\mu}\nonumber\\
\delta (F_\mu \phi) & = & -\xi^\lambda\partial_\lambda
(F_\mu\phi)
-\partial_\mu\xi^\lambda (F_\lambda\phi) - (1+i)x^\nu\partial_\mu b_\nu \label{trans1}
\end{eqnarray}
Thus both $F_\mu\phi$ and ${\bar{D}}_\mu\phi$ transform as four vectors under local Poincare transformations.

\hskip.5cm  The transformations of the second derivatives, however, do not have any Galileon contribution (see the last two  eq. of (\ref{pg1}). Consequently,
the second derivative $\partial_\mu\partial_\nu\phi$
should be replaced by 
$\nabla_a{\bar{\nabla}}_b\phi$ \footnote{A general discussion of localising the Poincare sector of a higher derivative theory is given in \cite{M1}.}. Explicit calculation shows that under local Poincare plus Galileon transformations $\nabla_a{\bar{\nabla}}_b
\phi$ transform in the same way as $\partial_\mu\partial_\nu\phi$ i.e.
\begin{equation}
\delta(\nabla_a {\bar{\nabla}}_b\phi)=-\xi^d \partial_d(\nabla_a {\bar{\nabla}}_b\phi)+{\theta_a}^d (\nabla_d {\bar{\nabla}}_b\phi)+{\theta_b}^d (\nabla_a {\bar{\nabla}}_d\phi)
\end{equation}
provided,
\begin{equation}
\nabla_a b_c = 0\label{cond2}
\end{equation}
This is a natural generalization of the condition $\partial_a b_c = 0$ in flat space. 
 We can define  $F_a$ as,
\begin{equation}
F_a={\Sigma_a}^{\mu}F_{\mu}
\end{equation}
and,
\begin{equation}
\nabla_a (F_b\phi)={\Sigma_a}^{\mu}[\partial_{\mu}({\Sigma_b}^{\nu}F_{\nu}\phi)+\frac{1}{2} M_{\mu}{}^{cd}\sigma_{cd}{\Sigma_b}^{\nu}F_{\nu}\phi]
\end{equation} 

for zero torsion, 
\begin{eqnarray}
\nabla_a{\Sigma_c}^\mu - \nabla_c{\Sigma_a}^\mu = 0\label{torsion}
\end{eqnarray}

thus we find,
\begin{equation}
\nabla_a (F_b\phi)-\nabla_b (F_a\phi)={\Sigma_a}^{\mu}{\Sigma_b}^{\nu}[\partial_{\mu}(F_{\nu}\phi)-\partial_{\nu}(F_{\mu}\phi)]\label{f}
\end{equation}

The variation under Galileon transformations now yields,
\begin{align}
\delta[\nabla_a (F_b\phi)-\nabla_b (F_a\phi)]&=(1+i){\Sigma_a}^{\mu}{\Sigma_b}^{\nu}(\partial_{\mu}b_{\nu}-\partial_{\nu}b_{\mu})
\notag\\&=(1+i)(\nabla_a b_b-\nabla_b b_a)=0
\end{align}
on account of (\ref{cond2}). This implies that the choice, 
\begin{equation}
[\nabla_a (F_b\phi)-\nabla_b (F_a\phi)]=0 \label{x}
\end{equation}
may be consistently implemented. 

Now  the commutator of ${\bar{\nabla}}_a$ and ${\bar{\nabla}}_b$ are given by

  \begin{eqnarray}
 [ {\bar{\nabla}}_a , {\bar{\nabla}}_b]\phi = [ {\nabla}_a , {\nabla}_b]\phi +[ {\nabla}_a , F_b]\phi +[F_a , {\nabla}_b]\phi +[F_{a},F_{b})]\phi
\end{eqnarray}
 
 After some calculation , using the symmetric property of $P_{\mu\nu}$ and the condition (\ref{x}) we get
 
  \begin{eqnarray}
 [ {\bar{\nabla}}_a , {\bar{\nabla}}_b]\phi &=&  \Sigma_{a}{}^{\mu}\Sigma_{b}{}^{\nu} \bigg[ \frac{1}{2}\sigma_{\alpha\beta} \bigg( \partial_{\mu}M_{\nu}{}^{\alpha\beta}-\partial_{\nu}M_{\mu}{}^{\alpha\beta}+M^{\alpha}{}_{\rho\mu} M^{\rho\beta}{}_{\nu}-M^{\alpha}{}_{\rho\nu} M^{\rho\beta}{}_{\mu}\bigg) \nonumber\\          &  &\qquad \qquad + (1+i)x^{\rho}(\partial_{\mu}P_{\nu\rho}-\partial_{\nu}P_{\mu\rho})\nonumber\\          &  &\qquad \qquad + \frac{1}{2}(1+i)x^{\rho}\sigma_{cd}(P_{\mu\rho}M_{\nu}{}^{cd}-P_{\nu\rho}M_{\mu}{}^{cd})\bigg]\phi \label{212}
\end{eqnarray}
    

 compearing this with (\ref{cova1}) and using (\ref{rici}) we get
 
 \begin{eqnarray}
 R_{\mu\nu}{}^{\lambda\sigma} = \Sigma^{\lambda}{}_{\alpha}\Sigma^{\sigma}{}_{\beta}  R_{\mu\nu}{}^{\alpha \beta} &= & \Sigma^{\lambda}{}_{\alpha}\Sigma^{\sigma}{}_{\beta}\bigg[ ( \partial_{\mu}M_{\nu}{}^{\alpha \beta}-\partial_{\nu}M_{\mu}{}^{\alpha \beta}+M^{\alpha }{}_{\rho\mu} M^{\rho  \beta}{}_{\nu}-M^{\alpha }{}_{\rho\nu} M^{\rho  \beta}{}_{\mu}) \nonumber\\          &  &\qquad \qquad + 2(1+i)x^{\rho}\sigma^{\alpha \beta}(\partial_{\mu}P_{\nu\rho}-\partial_{\nu}P_{\mu\rho})\nonumber\\          &  &\qquad \qquad + (1+i)x^{\rho}(P_{\mu\rho}M_{\nu}{}^{\alpha \beta}-P_{\nu\rho}M_{\mu}{}^{\alpha \beta})\bigg]
 \end{eqnarray}
 Now from (\ref{Ri}) we get the Ricci tensor as,

\begin{eqnarray}
               R_{\nu \sigma} = R_{\mu\nu}{} ^{\mu}{}_{\sigma} &= & \Sigma^{\mu}{}_{\alpha}\Sigma_{\sigma\beta}\bigg[ ( \partial_{\mu}M_{\nu}{}^{\alpha \beta}-\partial_{\nu}M_{\mu}{}^{\alpha \beta}+M^{\alpha }{}_{\rho\mu} M^{\rho  \beta}{}_{\nu}-M^{\alpha }{}_{\rho\nu} M^{\rho  \beta}{}_{\mu}) \nonumber\\          &  &\qquad \qquad + 2(1+i)x^{\rho}\sigma^{\alpha \beta}(\partial_{\mu}P_{\nu\rho}-\partial_{\nu}P_{\mu\rho})\nonumber\\          &  &\qquad \qquad + (1+i)x^{\rho}(P_{\mu\rho}M_{\nu}{}^{\alpha \beta}-P_{\nu\rho}M_{\mu}{}^{\alpha \beta})\bigg]
\label{gobbor} 
 \end{eqnarray}
 The equation (\ref{gobbor}) is to say the least, {\bf{extraordinary}}. The implication of the different terms are easy to recognized . In this connection it may be noted that the fracton action (\ref{1}) does not contain any mass term still it is interacting with gravity . Hence we can conclude that the gravity of fractons is an emergent one. The second point is obtained from the second term  which shows that there is a self interaction of the fractons as we see if the external gravity which is contain in the first and the third term put to zero there is still a non zero curvature . We also see to our satisfaction that these interactions have nothing to do with the symmetric tensor gauge field $B_{\mu\nu}$ introduced in connection with the electromagnetic interaction. The gravitational interaction of the fracton seems to be a deep concept which requires further investigations.

 \section{Conclusion :}
 In this paper  we have provided a co-varinat and gauge invarient version of the fracton gauge theory . This construction is built on the existing fracton gauge theory proposed in \cite{pret} . The seminal concepts of the gauge principle and the importance of matter Lagrangian are the two premises on which the fracton gauge theory \cite{pret} is universally viewed and ours is not an exception .   However, so far the question of  space-time symmetries were not much emphasized . In fact model mooted in  \cite{pret} is reported to have no boost in-variance. But the boost invariant is related to the equivalence of inertial frame, a concept which is so dear in physics that this point must be re-examine . As far as we have followed the literature of fracton gauge theory   we don't see any major analysis of this issue , but a gauge invariant theory in a particular inertial frame of reference may break down in another boosted frame. We have emphasized the issue of co-variance and gauge in-variance simultaneously in this paper.

 \hskip.5cm  Another very important issue is  that the fracton matter must  be represented by a Lagrangian field theory which has two conserved current . The physical interaction here is  the  electromagnetic interaction which is a one parameter Abelian gauge theory , will allow only one gauge degree of freedom. It is surprising that this question was never in the foreground . A related question is how to construct the Lagrangian of a free fracton matter with two  global gauge symmetries . Answer to this question is hard to get with in the premises  of the existing fracton gauge theory . We have shown in this paper that this comes quite naturally if we remember that the number of gauge parameters may be less than the number of conservation laws for a higher order Lagrangian theory \cite{pmb}\cite{pb}. The accompanying theory is essentially unstable as the equations of motion  are of higher order. It is well known that in general the higher order theories are infested with ostrogradsky instabilities. The way-out to this problem has been worked out in this paper by the use of a very special kind of scalar field which in-spite of being a higher order theory has usual second order equation of motion. The example of such scalar fields came from the research on modified gravity! This is the Galileon matter , the properties of which were determine \cite{nic} from a certain limit of DGP model \cite{dva}. Though Galileons are a hotbed  of intense research  all those are in connection with gravity theory , the use which we made here has no precedent \footnote{That is not to say Galileonic symmetries were unknown in this field . However the point is no body proposed that the Galileons are the source of fracton matter.}.

 \hskip .5cm    We have postulated  a complex scalar field consisting of the Galileon-like scalar \cite{nic},  so that the theory may have a hitherto unknown  complex shift symmetry . This theory is potent enough in explaining many crucial aspects of the fracton phase of matter which is demonstrated here. Among this the origin of two conservation laws , appearance of the tensor gauge theory , the physical content of the``generalized Maxwell's theory"     , the gravitational interaction of the fractons are included. All these are possible due to the  availability of  the co-variant and gauge in-variant model established in this paper .  Also the complex scalar field appearing here is a unique construction which may have significant role in other front ranking problems  of theoretical physics .\footnote{ For instance, during the writing of the paper we are informed that 
in connection with an alternative suggestion to quark confinement, the term fracton was first used \cite{khlopov}. Apparently, the symmetries of their problem has remarkable similarities with ours. Conclusions based on symmetry are often seen to be realized. }

\hskip .5cm  In short  this paper introduces a new way of looking at the fracton gauge theory which may very important in future research.

\section{Acknowledgments}

Sk. moinuddin would like to thank CSIR India for the fellowship provided to him.(File no: 08/606(0005)/2019-EMR-I). 

 Both SM and PM thanks R. Banerjee for useful comments.


\end{document}